\documentclass[twocolumn,aps,prd,superscriptaddress,nofootinbib,floatfix]{revtex4}

\usepackage{graphicx,multirow}
\usepackage[hypertex]{hyperref}

\newcommand{\nn}{\nonumber}
\newcommand{\beq}{\begin{equation}}
\newcommand{\eeq}{\end{equation}}
\newcommand{\bea}{\begin{eqnarray}}
\newcommand{\eea}{\end{eqnarray}}
\newcommand{\TeV}{\text{ TeV}}
\newcommand{\GeV}{\text{ GeV}}

\newcommand{\qqbar}{\ensuremath{{q\bar q}}}

\newcommand{\ttbar}{\ensuremath{{t\bar t}}}
\newcommand{\mttbar}{\ensuremath{m_{t\bar t}}}
\def\d{{\rm d}}

\makeatletter
\def\simgt{\mathrel{\lower2.5pt\vbox{\lineskip=0pt\baselineskip=0pt
           \hbox{$>$}\hbox{$\sim$}}}}
\def\simlt{\mathrel{\lower2.5pt\vbox{\lineskip=0pt\baselineskip=0pt
           \hbox{$<$}\hbox{$\sim$}}}}
\makeatother

\begin{document}

\title{\boldmath Comment on measuring the \ttbar\ forward-backward asymmetry at
ATLAS and CMS}

\author{Jean-Fran\c{c}ois Arguin}
\affiliation{Ernest Orlando Lawrence Berkeley National Laboratory,
University of California, Berkeley, CA 94720}

\author{Marat Freytsis}
\affiliation{Ernest Orlando Lawrence Berkeley National Laboratory,
University of California, Berkeley, CA 94720}
\affiliation{Berkeley Center for Theoretical Physics, Department of Physics,
University of California, Berkeley, CA 94720}

\author{Zoltan Ligeti}
\affiliation{Ernest Orlando Lawrence Berkeley National Laboratory,
University of California, Berkeley, CA 94720}

\begin{abstract}

We suggest a new possibility for ATLAS and CMS to explore the \ttbar\
forward-backward asymmetry measured at the Tevatron, by attempting to
reconstruct \ttbar\ events, with one of the tops decaying semileptonically in
the central region ($|\eta| < 2.5$) and the other decaying hadronically in the
forward region ($|\eta| > 2.5$).  For several models which give comparable
Tevatron signals, we study the charge asymmetry at the LHC as a function of cuts
on $|\eta|$ and on the \ttbar\ invariant mass, \mttbar.  We show that there is
an interesting  complementarity between cuts on $|\eta|$ and \mttbar\ to
suppress the dominant and symmetric $gg \to t\bar t$ rate, and different
combinations of cuts enhance the distinguishing power between models.  This
complementarity is likely to hold in other new physics scenarios as well, which
affect the \ttbar\ cross section, so it motivates extending \ttbar\
reconstruction to higher $|\eta|$.

\end{abstract}

\maketitle

\section{Introduction}

The hints of an enhancement of the forward-backward asymmetry in the production
of \ttbar\ pairs seen by CDF~\cite{Aaltonen:2008hc} and D\O~\cite{Abazov:2007qb}
a few years ago generated a lot of interest.  The top quark, via its large
coupling to the Higgs, can be especially sensitive to new physics at the TeV
scale.  The hint of a possible deviation from the standard model has become more
significant recently, when CDF found that the asymmetry arises from \ttbar\
events with high invariant masses~\cite{Aaltonen:2011kc},
\bea\label{CDFasym}
A_\ttbar(m_\ttbar > 450\,{\rm GeV}) &=& 0.475 \pm 0.114 \,, \nn \\ 
A_\ttbar(m_\ttbar < 450\,{\rm GeV}) &=& -0.116 \pm 0.153 \,,
\eea
while the standard model predictions are $0.088 \pm 0.013$ and $0.040\pm
0.006$, respectively.
The D\O\ result, only available integrated over \mttbar, and uncorrected for
effects from reconstruction or selection, $A_\ttbar = 0.08 \pm
0.04$~\cite{D0afb}, is consistent with the integrated CDF result, $A_\ttbar =
0.158 \pm 0.075$~\cite{Aaltonen:2011kc}.  So is the CDF measurement in the
dilepton channel, $A_\ttbar = 0.417 \pm 0.157$~\cite{CDFdilepton}. The large
asymmetry at high masses points toward tree-level exchange of a new particle,
with large couplings to first and third generation quarks~\cite{Jung:2009jz,
Frampton:2009rk, Shu:2009xf, Arhrib:2009hu, Degrande:2010kt, Cheung:2011qa,
Bai:2011ed, Gresham:2011dg, Grinstein:2011yv, Ligeti:2011vt, Gresham:2011pa}.

The identically defined forward-backward asymmetry vanishes at the LHC, since it
is a symmetric collider.  It has long been known that the same physics would
manifest itself in a different pseudorapidity distribution for the $t$ and the
$\bar t$ quarks~\cite{Kuhn:1998kw}, yielding a nonzero charge asymmetry,
\beq\label{Acdef}
A_c = \frac{N\big(|\eta_t|>|\eta_{\bar t}|\big) - N\big(|\eta_{\bar t}|>|\eta_t|\big)}
  {N\big(|\eta_t|>|\eta_{\bar t}|\big) + N\big(|\eta_{\bar t}|>|\eta_t|\big)} \,.
\eeq
This asymmetry is interesting to study for cuts that are symmetric between the
$t$ and $\bar t$ decays.  However, there is a severe dilution at the LHC due to
the dominance of the $gg\to\ttbar$ process, which generates no asymmetry, over
$\qqbar \to \ttbar$, which dominates at the Tevatron.  A measurement of this
asymmetry has  been performed by the CMS collaboration using $\int\!
\mathcal{L}\, \d t = 1.09\ \mathrm{fb}^{-1}$ data, integrated over $\mttbar$,
yielding $A_c = -0.019\pm0.030\mathrm{(stat.)}\pm 0.019
\mathrm{(syst.)}$~\cite{CMSAfb}. A similar measurement, with the asymmetry
defined for top rapidities instead of pseudorapidities, using  $0.70\
\mathrm{fb}^{-1}$ of data by the ATLAS collaboration yields $A'_c =
-0.024\pm0.016\mathrm{(stat.)}\pm 0.023 \mathrm{(syst.)}$~\cite{ATLASAfb}. The
statistical uncertainty of such measurements are expected to approach the
percent level with a few inverse femtobarns, but the systematic uncertainties
are not expected to improve at the same pace. Assuming that systematic
uncertainties better than a couple of percent will be difficult to achieve in
the near future, we attempt to maximize the size of $A_c$  to a level which
would allow to observe a significant excess by selecting a favorable phase space
of $\ttbar$ events, even at the price of reducing the $\ttbar$ sample, given the
large LHC integrated luminosity expected in the near future.

The role of the $\qqbar \to \ttbar$ process can be enhanced by demanding a high
invariant mass~\cite{Bai:2011ed, Strassler:2011vr, AguilarSaavedra:2011hz}, and
the models proposed to fit the CDF measurement generically predict at least a
factor of 2\,--\,3 enhancement of $\d\sigma / \d\mttbar$ at the LHC at \mttbar\
above 1\,--\,1.5~TeV~\cite{Ligeti:2011vt, Delaunay:2011gv,
AguilarSaavedra:2011vw}.  However, in the SM only about 1\% [0.1\%] of the
\ttbar\ events pass the $\mttbar > 1$~TeV [1.5~TeV] cut. These models also
predict an ${\cal O}(10\%)$ enhancement of the inclusive \ttbar\ cross section
at the LHC. Both of these effects would be hard to detect with large
significance in the near future, due to either low statistics or theoretical and
experimental uncertainties. 

Another possibility to enhance the $\qqbar \to \ttbar$ process and increase the
sensitivity to the charge asymmetry is to select events boosted in the beam
direction (see, e.g., Ref.~\cite{Wang:2010du,Bai:2011uk}).  A proposal to carry
out such an analysis at LHCb was made in Ref.~\cite{Kagan:2011yx} where only the
muon and $b$-jet of a leptonically decaying top are identified in the detector
range of  $2 < |\eta| < 5$. The $\ttbar$ physics programs at CMS and ATLAS have
so far concentrated on selecting leptons and jets within $|\eta| \lesssim 2.5$
motivated by the availability of the full detector features in that range, such
as lepton identification, requiring the lepton to pass through the tracking
system and either the fine-grained electromagnetic calorimeter (for electrons)
or muon  spectrometer (for muons), and $b$-tagging, again requiring tracking.
However, the calorimeters of both experiments have the capability to identify
and measure jets up to $|\eta| \lesssim 4.5$. Several physics measurements have
already been performed using such forward jets, prominent examples being the
observation of single top quark production by CMS~\cite{CMSsingletop} and
ATLAS~\cite{ATLASsingletop}. Therefore, in this paper, we assume that the
reconstruction of $\ttbar$ events can be extended to using forward jets in an
attempt to increase the observed $\ttbar$ asymmetry. We note that some of the
experimental difficulties involved by increasing the jet $|\eta|$ range cannot
be covered in this study, such as the increased jet energy scale uncertainty.
These questions will require a dedicated study using the full ATLAS and CMS
detector simulations to be answered properly.

\section{Models}

To explore the dependence of the charge asymmetry signal on $\eta$ and \mttbar,
we consider several models recently studied in the literature.  Their parameter
values are chosen such that they give rise to roughly comparable
$A_\ttbar(m_\ttbar > 450\,{\rm GeV})$ signals at the Tevatron.

Our benchmark models are (i) a $Z'$ coupling off-diagonally to
$ut$~\cite{Jung:2009jz}; (ii) an axigluon coupling with opposite sign to first
and third generation quarks such that $g_A^u = -g_A^t$~\cite{Frampton:2009rk};
(iii) and a scalar triplet diquark (the ``high-mass'' reference point in
Ref.~\cite{Ligeti:2011vt}). The masses and couplings of these models are
\bea
Z':  &&  m_{Z'} = 260 \GeV, \qquad \alpha_{Z'} = 0.048\,, \nn\\
\mbox{Axigluon\,:}  &&  m_A = 2 \TeV, \qquad\quad~~\, g_A = 2.4\,, \nn\\
\mbox{Scalar {\boldmath $3$}\,:} && m_S = 750 \GeV, \qquad\quad~ 
  \lambda = 3.0\,. 
\eea
The contributions to the Tevatron forward-backward asymmetries in these models
are shown in Table~\ref{tab:models}, which also shows the relative change in the
\ttbar\ cross sections at the Tevatron and at the LHC due to these new physics
contributions.  We chose these models because (as emphasized in
Ref.~\cite{Gresham:2011pa}) due to the different angular distributions in vector
and scalar exchanges in different channels ($s$, $t$, or $u$) for the new
particles, we expect to get a good sampling of the dependence of new physics
scenarios on the selection criteria we study. (Other models may in fact provide
better fits to the latest data; see, e.g., Refs.~\cite{Barcelo:2011vk,
Tavares:2011zg}.)

\begin{table}[tb]
\begin{tabular}{c||c|c|c}
\hline\hline
\multirow{2}{*}{Predictions}  & \multicolumn{3}{c}{new physics models} \\ 
\cline{2-4}
  &  $\quad Z'\quad$  &  ~Axigluon~  &  ~Scalar {\boldmath $3$}~  \\
\hline\hline
$A_\ttbar^{\rm TEV}(m_\ttbar > 450\,{\rm GeV})$  &
  $0.30$  &  $0.26$  &  $0.29$ \\ 
$A_\ttbar^{\rm TEV}$  &
  $0.15$  &  $0.14$  &  $0.17$ \\
\hline
$\sigma_\ttbar^{\rm TEV}/\sigma_\ttbar^{\rm TEV,SM}$
  &  $0.85$  &  $1.08$  &  $1.19$  \\ 
$\sigma_\ttbar^{\rm LHC}/\sigma_\ttbar^{\rm LHC,SM}$
  &  $1.01$  &  $1.16$  &  $1.11$ \\ 
\hline
\end{tabular}
\caption{Comparison of the three models studied.  The prediction for the
experimentally relevant measurable asymmetries is given by adding to the
numbers in the first [second] row the SM contributions, $A_{\ttbar}^{\rm
SM}=0.09$ [$0.04$]~\cite{Aaltonen:2011kc}.}
\label{tab:models}
\end{table}

In Table~\ref{tab:models}, we include for the Tevatron calculations an $|\eta| <
2.0$ cut, while for the LHC we impose $|\eta| < 2.5$.  The asymmetry predictions
for the new physics models are quoted from the leading order calculation (both
in Table~\ref{tab:models} and thereafter), so the SM contribution, which starts
at next-to-leading order, should be added to obtain the experimental
predictions.  To calculate the effect of the cuts on the charge asymmetry
generated in the SM (starting at next-to-leading order) we used
MCFM~\cite{Campbell:2010ff}, while for the NP contributions (at leading order)
we used MadGraph~\cite{Alwall:2011uj}.  There is an additional electroweak
contribution to the asymmetry~\cite{Chiu:2008vv,Hollik:2011ps} in the SM, which
has neither been included in our results nor in any of the numerical predictions
in the literature to our knowledge.  The related uncertainty is small, so it
should not affect our conclusions about the observability of a deviation from
the SM as a function of the different cuts.

\section{Cuts and Results}

We study the effect of imposing combinations of cuts on the reconstructed top
quarks' pseudorapidities ($R_i$) and invariant masses ($M_i$) in \ttbar\ events,
to enhance the charge asymmetry signal, as well as the effect on the signal
efficiency $\varepsilon$, defined as $\sigma_{\text{cuts}}/\sigma_{t\bar{t}}$.
We study the~cuts
\bea\label{cuts}
&& R_1:\quad |\eta_{1,2}| < 2.5\,, \nn\\*
&& R_2:\quad |\eta_1| < 2.5 \mbox{~~and~~} |\eta_2| < 4.5\,, \nn\\*
&& R_3:\quad |\eta_1| < 2.5 \mbox{~~and~~} 2.5 < |\eta_2| < 4.5\,, \nn\\*
&& M_1:\quad \mttbar > 450\,\GeV\,, \nn\\*
&& M_2:\quad \mttbar > 550\,\GeV\,,
\eea
motivated by the ATLAS and CMS detector geometries and the CDF analysis. 
Table~\ref{tab:ttbarcuts} shows the efficiencies, $\varepsilon$, and the
generated charge asymmetries, $A_c$, for the SM and our three benchmark models
in the hypothetical scenario in which the cuts in Eq.~(\ref{cuts}) could be
imposed on the $t$ and the $\bar t$ directly.  Table~\ref{tab:partoncuts} shows
the more realistic and experimentally more relevant results, when the cuts are
imposed on the observable decay products of the $t$ and the $\bar t$ instead. 
For the semileptonically decaying top in the central region we only impose
$|\eta| < 2.5$ on the lepton and the $b$-jet (and not on the neutrino).  By
the cuts $R_1$ and $R_2$ we now mean that each of the three jets from the
hadronically decaying top quark must satisfy $|\eta| < 2.5$ and $|\eta| < 4.5$,
respectively, but we impose no constraint on the pseudorapidity of the
reconstructed top itself.  For $R_3$ we keep the $|\eta| < 4.5$ selection
criteria on the jets of the decay products of the hadronically decaying top
quark, but we demand that the reconstructed top quark has $|\eta| > 2.5$. Note
that we do not impose an upper bound on the reconstructed pseudorapidity of the 
hadronically decaying top, thus including some tops at $|\eta| > 4.5$, with
decay products distributed across the beam axis from the direction of the top.

In addition to these cuts, to obtain the results in Table~\ref{tab:partoncuts},
we also imposed
\bea\label{MGcuts}
p_T^{\rm jet} &>& 25\,\GeV\,, \nn\\*
p_T^\ell &>& 20\,\GeV\,, \nn\\*
E_T^{\rm \nu} &>& 25 \GeV\,,
\eea
following similar selections in the ATLAS and CMS analyses.
We expect the leading order MadGraph calculation to give a useful guide to
the expected asymmetries and efficiencies in Tables~\ref{tab:ttbarcuts} and
\ref{tab:partoncuts} as a function of the kinematical cuts.
As mentioned above, the predicted asymmetry is the sum of the SM and NP
contributions in Tables~\ref{tab:ttbarcuts} and \ref{tab:partoncuts}, to a very
good approximation. 

\begin{figure}[t]
\includegraphics[width=\columnwidth]{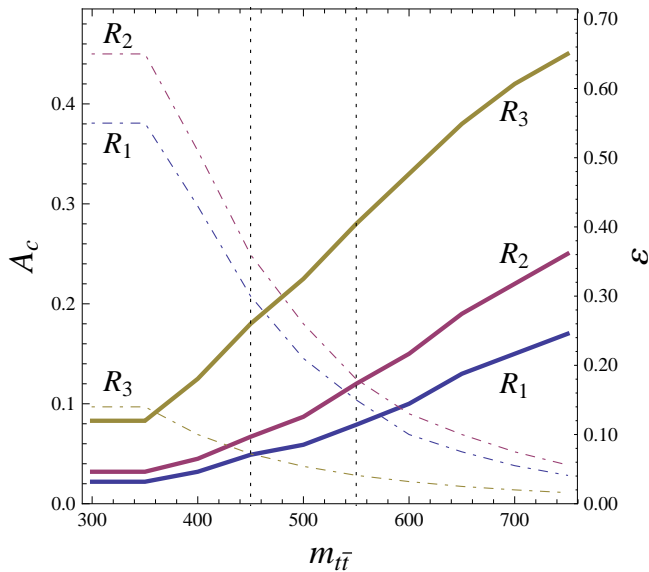}
\caption{Asymmetries (solid curves, legend on the left) and efficiencies (dashed
curves, legend on the right) for the $Z'$ model as a function of the lower cut
on $m_{t\bar{t}}$. The curves are labelled as showing the effect of the mass
cuts for $R_1$, $R_2$, and $R_3$ acceptances, respectively, and correspond to
decayed tops, as in Table~\ref{tab:partoncuts}. Vertical dotted lines indicate
the mass cuts listed in Table~\ref{tab:partoncuts}, with asymmetries/efficiencies
for the cuts below $m_{t\bar{t}} = 2m_t$ remaining constant.}
\label{fig:Zpttasym}
\end{figure}

\begin{figure}[t]
\includegraphics[width=\columnwidth]{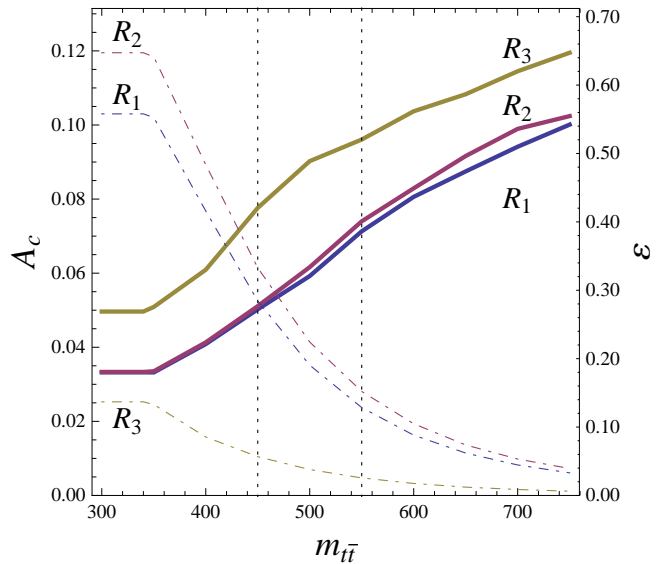}
\caption{Asymmetries and efficiencies for the axigluon model, as a function of
$m_{t\bar{t}}$. The notations are as in Fig.~\ref{fig:Zpttasym}.}
\label{fig:Apttasym}
\end{figure}

\begin{figure}[t]
\includegraphics[width=\columnwidth]{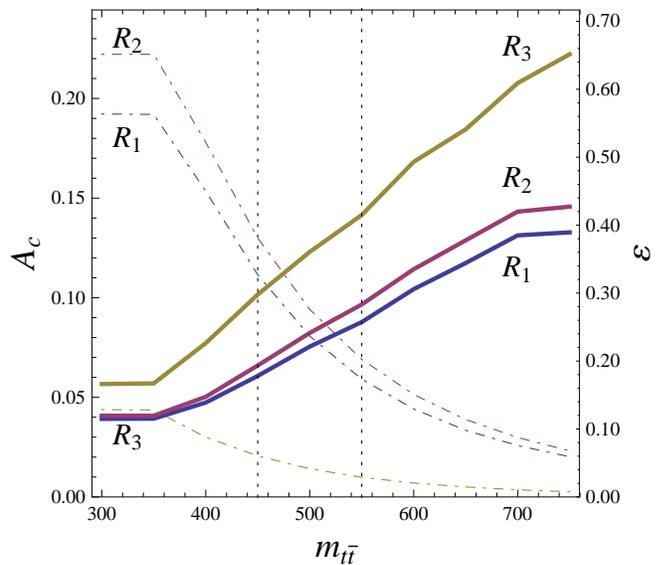}
\caption{Asymmetries and efficiencies for the scalar triplet model, as a
function of $m_{t\bar{t}}$. The notations are as in Fig.~\ref{fig:Zpttasym}.}
\label{fig:Tpttasym}
\end{figure}

In Figures~\ref{fig:Zpttasym}, \ref{fig:Apttasym}, and \ref{fig:Tpttasym}, we
plot the charge asymmetries and efficiencies in the three benchmark models, as
functions of the lower cut on \mttbar.  For no cut (that is, \mttbar\ lower cut
below $2m_t$), the cut at 450~GeV, and at 550~GeV, the plotted values reproduce
those in Table~\ref{tab:partoncuts}.

\begin{table}[t]
\begin{tabular}{c||c|c|c|c}
\hline\hline
\multirow{2}{*}{Cuts}  &  SM  & \multicolumn{3}{c}{new physics models} \\ 
\cline{3-5}
  &  MCFM  &  $Z'$  &  Axigluon  &  Scalar {\boldmath $3$}  \\
\hline\hline
\multirow{2}{*}{$R_1$}  &  \multirow{2}{*}{$A_c = 0.011$}
  &  $A_c = 0.019$  &  $A_c = 0.025$  &  $A_c = 0.038$ \\ 
  &  &  $\varepsilon = 0.77$  &  $\varepsilon = 0.78$ & $\varepsilon = 0.79$ \\ 
\hline
\multirow{2}{*}{$R_2$}  &  \multirow{2}{*}{$A_c = 0.018$}
  &  $A_c = 0.034$  &  $A_c = 0.031$  &  $A_c = 0.044$ \\ 
  &  &  $\varepsilon = 0.95$  &  $\varepsilon = 0.94$ &  $\varepsilon = 0.95$ \\ 
\hline
\multirow{2}{*}{$R_3$}  &  \multirow{2}{*}{$A_c = 0.028$}
  &  $A_c = 0.10$  &  $A_c = 0.058$  &  $A_c = 0.072$ \\ 
  &  &  $\varepsilon = 0.18$  &  $\varepsilon = 0.17$  &  $\varepsilon = 0.16$ \\ 
\hline\hline
\multirow{2}{*}{$R_1 \,\&\, M_1$}  &  \multirow{2}{*}{$A_c = 0.018$}
  &  $A_c = 0.038$  &  $A_c = 0.040$  &  $A_c = 0.059$ \\ 
  &  &  $\varepsilon = 0.44$  &  $\varepsilon = 0.42$  &  $\varepsilon = 0.48$ \\ 
\hline
\multirow{2}{*}{$R_2 \,\&\, M_1$}  &  \multirow{2}{*}{$A_c = 0.021$}
  &  $A_c = 0.064$  &  $A_c = 0.046$  &  $A_c = 0.068$  \\ 
  &  &  $\varepsilon = 0.54$  &  $\varepsilon = 0.50$  &  $\varepsilon = 0.57$ \\ 
\hline
\multirow{2}{*}{$R_3 \,\&\, M_1$}  &  \multirow{2}{*}{$A_c = 0.037$}
  &  $A_c = 0.18$  &  $A_c = 0.080$  &  $A_c = 0.12$ \\ 
  &  &  $\varepsilon = 0.10$  &  $\varepsilon = 0.082$  &  $\varepsilon = 0.087$ \\ 
\hline\hline
\multirow{2}{*}{$R_1 \,\&\, M_2$}  &  \multirow{2}{*}{$A_c = 0.022$}
  &  $A_c = 0.075$  &  $A_c = 0.061$  &  $A_c = 0.089$ \\ 
  &  &  $\varepsilon = 0.21$  &  $\varepsilon = 0.19$  &  $\varepsilon = 0.25$ \\ 
\hline
\multirow{2}{*}{$R_2 \,\&\, M_2$}  &  \multirow{2}{*}{$A_c = 0.029$}
  &  $A_c = 0.12$  &  $A_c = 0.10$  &  $A_c = 0.10$ \\ 
  &  &  $\varepsilon = 0.27$  &  $\varepsilon = 0.22$  &  $\varepsilon = 0.29$ \\ 
\hline
\multirow{2}{*}{$R_3 \,\&\, M_2$}  &  \multirow{2}{*}{$A_c = 0.041$}
  &  $A_c = 0.29$  &  $A_c = 0.10$  &  $A_c = 0.16$ \\ 
  &  &  $\varepsilon = 0.057$  &  $\varepsilon = 0.036$  &  $\varepsilon = 0.041$ \\ 
\hline\hline
\end{tabular}
\caption{Asymmetries and efficiencies resulting from the cuts defined in
Eq.~(\ref{cuts}), applied at the \ttbar\ level. The asymmetries for the NP
models are calculated including both the SM and NP at leading order, to which
the next-to-leading order SM results (in the second column) should be added to
obtain the predictions for the observable asymmetries.}
\label{tab:ttbarcuts}
\end{table}

\begin{table}[t]
\begin{tabular}{c||c|c|c|c}
\hline\hline
\multirow{2}{*}{Cuts}  &  SM  &  \multicolumn{3}{c}{new physics models} \\
\cline{3-5}
  &  MCFM  &  $Z'$  &  Axigluon  &  Scalar {\boldmath $3$}  \\
\hline\hline
\multirow{2}{*}{$R_1$}	&  \multirow{2}{*}{$A_c = 0.014$}
  &  $A_c = 0.022$  &  $A_c = 0.032$  &  $A_c = 0.041$ \\ 
  &  &  $\varepsilon = 0.55$  &  $\varepsilon = 0.56$  &  $\varepsilon = 0.56$ \\ 
\hline
\multirow{2}{*}{$R_2$}	&  \multirow{2}{*}{$A_c = 0.019$}
  &  $A_c = 0.032$  &  $A_c = 0.033$  &  $A_c = 0.042$  \\ 
  &  &  $\varepsilon = 0.65$  &  $\varepsilon = 0.65$ &  $\varepsilon = 0.65$  \\ 
\hline
\multirow{2}{*}{$R_3$}  &  \multirow{2}{*}{$A_c = 0.020$}
  &  $A_c = 0.083$  &  $A_c = 0.048$  &  $A_c = 0.054$  \\ 
  &  &  $\varepsilon = 0.14$  &  $\varepsilon = 0.14$ &  $\varepsilon = 0.13$ \\ 
\hline\hline
\multirow{2}{*}{$R_1 \,\&\, M_1$}  &  \multirow{2}{*}{$A_c = 0.022$}
  &  $A_c = 0.049$  &  $A_c = 0.050$  &  $A_c = 0.062$ \\ 
  &  &  $\varepsilon = 0.30$  &  $\varepsilon = 0.28$  &  $\varepsilon = 0.33$ \\ 
\hline
\multirow{2}{*}{$R_2 \,\&\, M_1$}  &  \multirow{2}{*}{$A_c = 0.023$}
  &  $A_c = 0.067$  &  $A_c = 0.051$ &  $A_c = 0.068$ \\ 
  &  &  $\varepsilon = 0.36$  &  $\varepsilon = 0.33$ &  $\varepsilon = 0.38$ \\ 
\hline
\multirow{2}{*}{$R_3 \,\&\, M_1$}  &  \multirow{2}{*}{$A_c = 0.042$}
  &  $A_c = 0.18$  &  $A_c = 0.077$ &  $A_c = 0.099$   \\ 
  &  &  $\varepsilon = 0.072$  &  $\varepsilon = 0.057$ &  $\varepsilon = 0.060$  \\ 
\hline\hline
\multirow{2}{*}{$R_1 \,\&\, M_2$}  &  \multirow{2}{*}{$A_c = 0.025$}
  &  $A_c = 0.079$  &  $A_c = 0.070$  &  $A_c = 0.092$ \\ 
  &  &  $\varepsilon = 0.15$  &  $\varepsilon = 0.13$  &  $\varepsilon = 0.17$ \\ 
\hline
\multirow{2}{*}{$R_2 \,\&\, M_2$}  &  \multirow{2}{*}{$A_c = 0.023$}
  &  $A_c = 0.12$  &  $A_c = 0.072$  &  $A_c = 0.10$ \\ 
  &  &  $\varepsilon = 0.18$  &  $\varepsilon = 0.15$ &  $\varepsilon = 0.20$ \\ 
\hline
\multirow{2}{*}{$R_3 \,\&\, M_2$}  &  \multirow{2}{*}{$A_c = 0.044$}
  &  $A_c = 0.28$  &  $A_c = 0.092$ &  $A_c = 0.14$ \\ 
  &  &  $\varepsilon = 0.041$  &  $\varepsilon = 0.026$ &  $\varepsilon = 0.029$ \\ 
\hline\hline
\end{tabular}
\caption{Asymmetries and efficiencies resulting from applying the cuts in
Eq.~(\ref{cuts}) to the partonic decay products of the \ttbar\ pair, and also
including the experimental cuts in Eq.~(\ref{MGcuts}). To get the total asymmetry,
asymmetries should be added in the manner of Table~\ref{tab:ttbarcuts} 
}
\label{tab:partoncuts}
\end{table}

To assess the prospects for observing a given model for a given set of cuts, we
need to make some assumptions on the achievable experimental uncertainties. 
From the CMS result~\cite{CMSAfb} we assume that systematic uncertainties on
$A_c$ below $\approx$\,0.02 will be difficult to achieve in the near future.
Therefore, we assume that, for a clear observation of the $\ttbar$ asymmetry, one
is likely to need to choose cuts that increase $A_c$ above 0.10 for a given
model. We assume that the systematic uncertainties are the same for every cut,
although in reality is should be somewhat larger at high jet $\eta$ and large
$\mttbar$. A more realistic assessment of systematic uncertainties will require
a dedicated study with full detector simulations. Furthermore, based on the CMS
analysis we assume that a statistical uncertainty of $\approx$\,0.13 is
achievable with a dataset of 36 pb$^{-1}$  for the cuts $R_1$, and that it will
scale with the integrated luminosity. Taking into account the different relative
acceptances between the cuts, this allows us to provide a rough estimate of how
much data will be needed to observe a given model. Again, a full detector
simulation will be needed to make a precise estimate of the expected statistical
uncertainty for given cuts.

Referring to Table~\ref{tab:partoncuts},  We note that simply using the basic set of
cuts, $R_1$, will not be sufficient to obtain a clear signal of the  asymmetry.
Allowing or requesting tops with high pseudorapidity (cuts $R_2$ and $R_3$)
increases  the asymmetry for each model, but not to a level likely to be soon
observable experimentally.  By selecting events with $\mttbar > 550~$GeV for the
basic set of cuts $R_1$ ($R_1$ \& $M_2$), one can increase $A_c$ to  about 0.08,
which should be enough to start seeing an evidence of a $\ttbar$ asymmetry in
the near future, but which will probably require experimental systematic
uncertainties to improve significantly to obtain a clear observation. By
allowing jets of the hadronic top in the forward region and  selecting events
with $\mttbar > 550~$GeV ($R_2$ \& $M_2$), one can increase $A_c$ to above 0.10
in the $Z'$ and scalar triplet models. The acceptance is reduced by a factor
$\approx$\,3 compared to the basic cuts $R_1$, which results in statistical
uncertainties of order 0.02 for a dataset of 5~fb$^{-1}$ using the assumptions
above. The asymmetry can be substantially increased by requesting that the
hadronically decaying top has $|\eta| > 2.5$, but by paying the price of a
significantly reduced acceptance.  This $R_3$ cut increases the asymmetries by a
factor of about 2 and 1.5 for the  $Z'$ and scalar triplet models, respectively.
For example, the $Z'$ model, for which $A_c$ approaches 0.3 for the cut $R_3$ \&
$M_2$, should be clearly observable for datasets of 5 fb$^{-1}$, for which a
statistical uncertainty of roughly 0.04 would be achievable with the assumptions
described above. The increase is also present but not as significant for the
axigluon model.

This by itself provides a strong motivation for studying the behavior of $A_c$
as a function of the top pseudorapidity. Moreover, especially in concert with
cuts on \mttbar, considering changes in $A_c$ for varying top pseudorapidity
acceptances allows for differentiation between the various proposed models from
their qualitatively different behavior. For example, the increase in $A_c$ for
the axigluon comes almost entirely from the \mttbar\ cuts, as can be seen from
the nearly degenerate $R_1$ and $R_2$ lines in Fig.~\ref{fig:Apttasym}. On the
other hand, for the $Z'$ and the triplet, both the loosening of geometric cuts
and the \mttbar\ cut help in significantly increasing the asymmetry, with the
geometric cuts being more significant for the $Z'$ model. Larger datasets will
be needed to unambiguously distinguish the various models, rather than merely
detect the asymmetry.  More complex information contained in the kinematic
distributions, such as spin correlations~\cite{Krohn:2011tw}, may also increase
the discriminating power between models.

One concern about requiring the $t$ or the $\bar{t}$ to be produced at higher
pseudorapidities is that the resulting decay products may be too highly boosted
to be effectively separated. Furthermore, a boosted top analysis might be
complicated by the fact that the detector granularity worsens in the forward
region of the ATLAS and CMS calorimeters. Imposing the condition that all three
jets from the top with $|\eta| > 2.5$ be separated by $R > 0.4$ [0.6], we find
that the fraction of events failing to meet this condition is below 11\% [25\%]
for the $R_3$ \& $M_2$ cuts, and is more typically ${\cal O}(5\%)$ [${\cal
O}(15\%)$]. If one only imposes the condition that at least two of the three
jets are separated by $R > 0.4$ [0.6], and assumes the feasibility of an
analysis that selects only two jets for the hadronically decaying top, with one
of them at high mass due to the merged hadronically decaying W boson, then the
fraction of events failing the cut is ${\cal O}(0.1\%)$ [${\cal O}(1\%)$].
Therefore, we conclude that the experimental challenges created by the increased
amount of boosted tops at high $\eta$ and $\mttbar$ are likely to be
surmountable.

We note that more complex analysis techniques, beyond our simple analysis, could
increase the significance of a putative signal. A fully systematic survey 
consisting of a larger number of geometric cuts and their combinations, which
has not been carried out herein, could be explored in order to optimize the
sensitivity to a charge asymmetry.

\section{conclusions}

In conclusion, we explored the sensitivity to a class of new physics models,
motivated by the Tevatron forward-backward asymmetry in \ttbar\ production, that
could be gained by ATLAS and CMS if the pseudorapidity ranges for \ttbar\
reconstruction could be extended.   We found that a combination of cuts based on
pseudorapidities and invariant masses of the reconstructed tops in \ttbar\
events greatly enhances the experimental sensitivity compared to just using one
of these cuts.  While it is clear that one of the tops that decays
semileptonically and thus tags the flavor has to be in the central region,
$|\eta| < 2.5$, our study shows that it would clearly be worthwhile to explore
if the other one could be reconstructed at higher $|\eta|$, in its decay to
jets.  From our main results in Table~\ref{tab:partoncuts} and in the figures,
one clearly sees that if the $R_3$ selection cut (or something similar) becomes
experimentally feasible at ATLAS and/or CMS, that would substantially enhance
the sensitivity to new physics, and allow better discrimination between
different models.

\begin{acknowledgments}

We thank Jernej Kamenik, Gilad Perez, and especially Martin Schmaltz for
helpful conversations, and Johan Alwall and Tim Tait for help with MadGraph. 
Z.L.\ thanks the Aspen Center for Physics for hospitality while parts of this
work were completed.
This work was supported in part by the Director, Office of Science, Office of
High Energy Physics of the U.S.\ Department of Energy under contract
DE-AC02-05CH11231.

\end{acknowledgments}

\end{document}